\documentclass[twocolumn,showpacs,preprintnumbers,amsmath,amssymb,nofootinbib,floatfix]{revtex4}

\usepackage[dvips]{graphicx}
\usepackage{dcolumn}
\usepackage{bm}
\usepackage{braket}

\def\Journal#1#2#3#4{{#1} {\bf #2}, #3 (#4)}
\def\AA{Astron. \& Astrophys.}

\def\APSS{Astrophys. and Space Sci.}

\def\CPC{Chin. Phys. C}
\def\EPJC{Eur. Phys. J. C}

\def\JCAP{JCAP}

\def\MolePhys{Mole. Phys.}

\def\PLB{{Phys. Lett.} B}

\def\PRL{Phys. Rev. Lett.}
\def\PRD{Phys. Rev. D}

\def\PR{Phys. Rev.}

\begin{document}


\title{Parafermionic dark mater}

\author{Teruyuki Kitabayashi}
\email{teruyuki@tokai-u.jp}

\author{Masaki Yasu\`{e}}%
\email{yasue@keyaki.cc.u-tokai.ac.jp}
\affiliation{\vspace{5mm}%
\sl Department of Physics, Tokai University,
4-1-1 Kitakaname, Hiratsuka, Kanagawa 259-1292, Japan
}

\date{\today}

\begin{abstract}
To discuss the possible contribution of parafermions to the dark matter abundance, we extend the Boltzmann equation for fermionic dark matter to include parafermions.  Parafermions can accommodate $r$ particles per quantum state $(2\le r<\infty)$, where the parafermion of order $r=1$ is identical to the ordinary fermion. It is found that the parafermionic dark matter can be more abundant than the fermionic dark matter in the present universe. 
\end{abstract}

\maketitle



\section{Introduction}
Understanding the nature of dark matter is one of the outstanding problems to be solved in the particle physics and cosmology \cite{DM_review}. There are a lots of candidates of dark matter in particle physics, such as WIMPs, axion, sterile neutrinos, and so on. Currently proposed all candidates of particle dark matter are classified as fermions or bosons. Fermions and bosons are characterize by the maximum occupation-number $r$. One particle per quantum state is allowed for one fermion ($r=1$), while an infinite number of particles per quantum state is allowed for one boson ($r=\infty$). On the other hand, other particles satisfying a nonstandard statistics, parafermi statistics \cite{Ohnuki1982}, have been discussed to find the possible realization in physics. It is called ``parafermion" that accommodates $r$ particles per quantum state $(2\le r<\infty)$. 

The quantum field theory of parafields satisfies the basic conditions in general quantum field theory such as cluster property, the existence of the unique vacuum, positive property of norm of physical state vector \cite{Ohnuki1982}. In spite of these theoretically accepted properties, the two limiting cases corresponding to the lower limit of $r$ as a fermionic field ($r=1$) and to the upper limit of $r$ as a bosonic field ($r=\infty$) are chosen in the current physics. However, we have experienced that a new discovery in physics sometimes arose from a theoretical possibility supported by the mathematical consistency. For parafields, the simple algebra to be shown in the next section supports the existence of parafermions (and parabosons). To study yet unknown properties of parafermi fields may give a clue to find answers to questions: ``Why do the extremes survive in nature?"  and ``Is there any privilege in physics for fermions and bosons to serve as building blocks of the Universe?"  For parafermi fields, it was pointed out that the parafermi fields of order 1 and 3 occupied a very privileged position in parafermi theory \cite{Ohnuki1968PR}.  Although parafermions have never been observed and there is yet no successful model including parafermions, we expect that the study of parafermion gives us some hints to reveal some of answers to these questions. This is our motivation to discuss possible realization of parafermions in physics.

Some problems, which may be inherent in the conventional statistics, can be solved considering generalized statistics \cite{Ebadi2013JCAP,Xiao2011PLB}. For an example, the modification of the entropy bounds of the Bose, Fermi statistics supplied by the gravitational stability condition is discussed \cite{Xiao2011PLB}.  The relationship between quantum gravity and non-standard statistics including infinite statistics \cite{Green1953PR,Doplicher1971ICMP,Greenberg1990PRL,Greenberg1991PRD} is also clarified.

In this paper, we would like parafermions to enjoy their physics at the initial stage of the expanding universe, where parafermions may serve as seeds of dark matter. To be realistic, we derive a simple-formed Boltzmann equation to estimate the relic abundance of parafermionic dark matter. As a result, we find that the parafermionic dark matter can be more abundant than the fermionic dark matter in the present universe. 

This paper is organized as follows. In Sec.\ref{section:parafermion}, some thermodynamical or statistical quantities, such as partition function, distribution function, energy density, etc. for parafermion are shown. In Sec.\ref{section:boltzmann_equation}, first, we review the usually used simple form of the Boltzmann equation to calculate relic abundance of fermionic dark matter (\ref{subsection:fermionic_dark_matter}). Second, we extend the Boltzmann equation to include parafermionic dark matter (\ref{subsection:parafermionic_dark_matter}). In Sec.\ref{section:relic_abundance}, we perform a numerical calculation to estimate the relic abundance of parafermionic dark matter. Sec.\ref{section:summary} is devoted to a summary.

\section{Parafermion\label{section:parafermion}}
The theoretical possibility of parafermions is readily seen from the well-known algebra of the creation and annihilation operators \cite{Ohnuki1982}, respectively, denoted by $a_+$ and $a_-$($=a^\dagger_+$), which are required to satisfy $\left[N, a_\pm\right] = \pm a_\pm$, where $N$ is the number operator. Furthermore, $N = \left[a_+, a_-\right]/2$ can be chosen as the simplest solution to the Jacobi identity for $a_\pm$ and $N$. The relations among $a_\pm$ and $N$ coincide with the algebra of $SO(3)$, which has three generators $J_{1,2,3}$ satisfying $\left[J_3, J_{\pm}\right]=\pm J_{\pm}$ and $\left[J_+, J_-\right]=2J_3$, where $J_{\pm} = J_1\pm iJ_2$.  It is readily understood that $a_\pm$ and $N$ can be identified with $a_\pm=J_\pm$ and $N=J_3$, which finally explain the spectrum of parafermions of order $r$ to be determined by $r=2\ell$ with $2\ell+1$ states for $\ell=0,1/2,1,3/2,\cdots$. 

A parafermion of order $r$ is described by $r+1$ states spanned by $\left| 0\right\rangle$, $\left| 1\right\rangle$ ,$\cdots$, $\left| r\right\rangle$, where $\left| n\right\rangle$ ($n\geqq 0$) is the eigenvector of $N$ for a given eigen value $n+N_0$ ($N_0 =-\ell$). A parafermion of order $r=1$ has two states $\left| 0\right\rangle$ and $\left| 1\right\rangle$, which describe the states of the ordinary fermion. Similarly, a parafermion with $r=2$ has three states $\left| 0\right\rangle$, $\left| 1\right\rangle$ and $\left| 2\right\rangle$ and is allowed to accommodate at most 2 particles per quantum state. Noticing that the relation of $\left[N, a_\pm\right] = \pm a_\pm$ is equivalent to $\left[a_-, [a_+, a_-] \right]$ = $\left\{a_-, \{a_-, a_+\}\right\}$ - $\left\{a_+, \{a_-, a_-\}\right\}$ = $2a_-$, we observe that it is, for example, satisfied by $\left\{a_-,a_+\right\}=1$ with $a^2_-=0$ for $r=1$ and by $\left\{a^2_-,a_+\right\}=2a_-$ and $a_-a_+a_-=2a_-$ with $a^3_-=0$ for $r=2$.

For a paraboson of order $r$, this analysis is, however, irrelevant because $N = \left[a_+, a_-\right]/2$ becomes a $c$-number for a paraboson of order $1$ identical to the ordinary boson. The right answer is to choose $N = \left\{a_+,a_-\right\}/2$ as the next simple solution to the Jacobi identity. After some manipulation, it turns out that the identification of $J_3=N/2$ and $J_\pm=a^2_\pm/2$ leads $J_{1,2,3}$ to the algebra of $SO(2,1)(\simeq Sp(2,\Re))$ instead of $SO(3)$.  Since $SO(2,1)$ is a non-compact group, $N$ has an infinite number of eigenvalues, which allows parabosons to have an infinite number of states per quantum state.

Let us discuss a statistical property of parafermions. The grand-canonical partition function for quantum parafermionic gas of the chemical potential $\mu$ with temperature $T$ is
\begin{eqnarray}
\Xi &=& \prod_i \sum_{n_i=0}^{r_i}  e^{-(E_i-\mu)n_i/T},
\label{Eq:Xi}
\end{eqnarray}
where $r_i$ denotes the maximum occupation-number corresponding to the energy $E_i$ \cite{Landsberg1963MolePhys, Aldrovandi1983APSS}. The distribution function for parafermions  
\begin{eqnarray}
f &=& \frac{\sum_{n=0}^{r} n e^{-(E-\mu)n/T}}{\sum_{n=0}^{r} e^{-(E-\mu)n/T}} \nonumber \\
 &=& \left. x\frac{d}{dx} \ln \sum_{n=0}^r x^n \right|_{x=e^{-(E-\mu)/T}},
\label{Eq:f_parafermi}
\end{eqnarray}
is derived from Eq.(\ref{Eq:Xi}). It is obvious that the distribution function for fermions $f_{\rm FD}$ is given by $r = 1$.  Furthermore, the limit of $r=\infty$ leads to the distribution function for bosons $f_{\rm BE}$. Namely, $f_{\rm FD}$ and $f_{\rm BE}$ are obtained as
\begin{eqnarray}
f_{\rm FD} &=& \left. f \right|_{r=1}\nonumber \\
      &=& \left. x \frac{d}{dx} \ln (1+ x)  \right|_{x=e^{-(E-\mu)/T}}\nonumber \\
      &=& \frac{1}{e^{(E-\mu)/T}+ 1},
\label{Eq:f_FD}
\end{eqnarray}
and
\begin{eqnarray}
f_{\rm BE} &=& \lim_{r\rightarrow\infty} f  \nonumber \\
      &=& \left. x\frac{d}{dx} \ln (1- x)^{-1}  \right|_{x=e^{-(E-\mu)/T}}\nonumber \\
      &=& \frac{1}{e^{(E-\mu)/T} - 1}.
\label{Eq:f_BE}
\end{eqnarray}
The number density $n_i$, energy density $\rho_i$, pressure $P_i$ and entropy density $s_i$ of a parafermionic species $i$ with internal degrees of freedom $g_i$ ($g_i=2$ \cite{Ohnuki1982}) are given by
\begin{eqnarray}
n_i(t) &=& \frac{g_i}{(2\pi)^3} \int f_i (\vec{p}_i) d^3p_i,  \\
\rho_i(t)&=& \frac{g_i}{(2\pi)^3} \int E(\vec{p}_i) f_i(\vec{p}_i) d^3p_i, \\
P_i(t)&=& \frac{g_i}{(2\pi)^3} \int \frac{|\vec{p}_i|^2}{3E} f_i(\vec{p}_i) d^3p_i,
\end{eqnarray}
and
\begin{eqnarray}
s_i = \frac{\rho_i + P_i -\mu_i n_i}{T_i},
\label{Eq:entropy_density}
\end{eqnarray}
where $E^2=|\vec{p}|^2 + m^2$ \cite{KolbTurner1990}. 

For the relativistic particles ($T_i \gg m_i$), in the isotropic universe, we have
\begin{eqnarray}
n_i^{\rm rela} &=& \frac{g_i}{2\pi^2}T^3 \int_x^\infty f(u) u^2du, \label{Eq:n_rela} \\
\rho_i^{\rm rela} &=& \frac{g_i}{2\pi^2} T^4 \int_x^\infty f(u) u^3du,\label{Eq:rho_rela}\\
P_i^{\rm rela} &=& \frac{g_i}{6\pi^2}T^4 \int_x^\infty f(u) u^3du,\label{Eq:P_rela}
\end{eqnarray}
where
\begin{eqnarray}
x=\frac{m}{T},  \quad u=\frac{E}{T}.
\end{eqnarray}
From Eqs.(\ref{Eq:entropy_density}) - (\ref{Eq:P_rela}), the total energy density $\rho^{\rm rela}$ and total entropy density $s^{\rm rela}$ of relativistic particles are obtained as
\begin{eqnarray}
\rho^{\rm rela} = \frac{\pi^2}{30}g_\ast T^4,  \quad s^{\rm rela}  = \frac{2\pi^2}{45}g_{\ast s} T^3,  \nonumber
\end{eqnarray}
where $g_\ast$ and $g_{\ast s}$ are defined by
\begin{eqnarray}
g_\ast &=& \sum_{\rm i=bosons} g_i \left( \frac{T_i}{T} \right)^4 + \frac{7}{8} \sum_{\rm i=fermions} g_i \left( \frac{T_i}{T} \right)^4 \nonumber \\
&& + \gamma_\ast \sum_{\rm i=parafermions} g_i \left( \frac{T_i}{T} \right)^4 , 
\label{Eq:gast}
\end{eqnarray}
and
\begin{eqnarray}
g_{\ast s} &=& \sum_{\rm i=bosons} g_i \left( \frac{T_i}{T} \right)^3 + \frac{7}{8} \sum_{\rm i=fermions} g_i \left( \frac{T_i}{T} \right)^3 \nonumber \\
&& + \gamma_{\ast s} \sum_{\rm i=parafermions} g_i \left( \frac{T_i}{T} \right)^3 , 
\end{eqnarray}
respectively, denote the relativistic effective degrees of freedom for the energy density and for the entropy density. The factor $7/8$ accounts for the difference between the Fermi and Bose statistics, and the factor 
\begin{eqnarray}
\gamma_\ast = \frac{\rho_{\rm i=parafermion}^{\rm rela}}{\rho_{\rm i=boson}^{\rm rela}} \end{eqnarray}
and
\begin{eqnarray}
\gamma_{\ast s} = \frac{s_{\rm i=parafermion}^{\rm rela}}{s_{\rm i=boson}^{\rm rela}}
= \frac{\rho_{\rm i=parafermion}^{\rm rela}}{\rho_{\rm i=boson}^{\rm rela}} 
=\gamma_\ast
\end{eqnarray}
respectively, account for the difference between the parafermi and Bose statistics.

\begin{figure}[t]
\begin{center}
\includegraphics{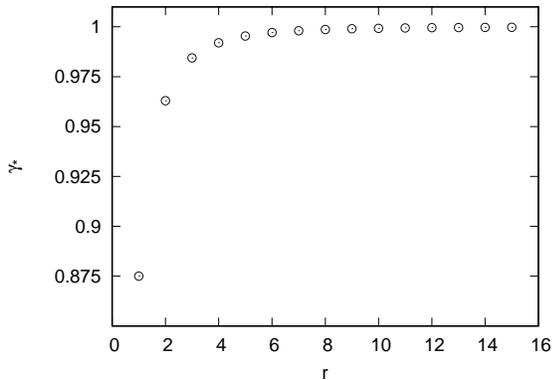}
\caption{The dependence of the statistical factor for parafermion $\gamma_\ast$ on the maximum occupation-number $r$. The relations of $\gamma_\ast |_{r=1}= 7/8$ and $\lim_{r \rightarrow \infty}\gamma_\ast = 1$ are satisfied as we expected.}
\label{fig:gamma}
\end{center}
\end{figure}

Since the parafermi statistics is regarded as the Fermi statistics in the case of $r=1$ and as the Bose statistics in the case of $r=\infty$, the following relation
\begin{eqnarray}
\gamma_\ast |_{r=1}= \frac{7}{8}, \quad 
\lim_{r \rightarrow \infty}\gamma_\ast = 1
\label{Eq:gamma_limit}
\end{eqnarray}
should be satisfied. FIG.\ref{fig:gamma} shows the dependence of the factor $\gamma_\ast$ on the maximum occupation-number $r$. The relation in Eq.(\ref{Eq:gamma_limit}) is satisfied as we expected.

\section{Boltzmann equation\label{section:boltzmann_equation}}
\subsection{Fermionic dark matter\label{subsection:fermionic_dark_matter}}
The time evolution of the phase space distribution functions as well as the number density for dark matter can be controlled by the Boltzmann equation \cite{KolbTurner1990}. Before we derive a Boltzmann equation for parafermionic dark matter, we review the usually used simple-formed Boltzmann equation for fermionic dark matter.

In the homogeneous and isotropic universe, the relativistic Boltzmann equation for $1+2 \leftrightarrow 3+4$ process is written in the form \cite{KolbTurner1990}
\begin{eqnarray}
\frac{dn_1}{dt} + 3Hn_1 = \frac{g_1}{(2\pi)^3} \int \frac{C[f]}{E_1}d^3p_1,
\end{eqnarray}
where $H=\pi T^2 M_{pl}^{-1}\sqrt{g_\ast/90}$ denotes the Hubble parameter and $C[f]$ denotes the collision term. The collision term evolves depending on the equation:
\begin{widetext}
\begin{eqnarray}
\frac{g_1}{(2\pi)^3} \int \frac{C[f]}{E_1}d^3p_1 &=& -\int d \Pi_1 d \Pi_2 d \Pi_3 d \Pi_4 (2\pi)^4 \delta^4(p_1+p_2-p_3-p_4) \\ 
&& \quad \times \left\{ \left| \mathcal{M}_{1+2 \rightarrow 3+4} \right|^2 f_1 f_2 (1 + \zeta_3 f_3)(1 + \zeta_4 f_4) - \left| \mathcal{M}_{3+4 \rightarrow 1+2} \right|^2 f_3 f_4 (1 + \zeta_1 f_1)(1 + \zeta_2 f_2) \right\} \nonumber
\end{eqnarray}
\end{widetext}
where
\begin{eqnarray}
\zeta_i = 
\begin{cases}
+1 &  ({\rm bosons})\\
-1 &  ({\rm fermions}) \\
\end{cases}
\end{eqnarray}
and
\begin{eqnarray}
d \Pi_i = \frac{g_i}{(2\pi)^3}\frac{d^3 p_i}{2E_i}.
\end{eqnarray}
The term of $(1 + \zeta_i f_i)$ shows the Pauli blocking of fermions ($\zeta_i=-1$) or the stimulated emission of bosons ($\zeta_i=+1$). 

We use a simpler version of the Boltzmann equation accepting the following assumptions that
\begin{enumerate}
\item CP invariance is satisfied.
\item The particles $1,2,3$ and $4$ are fermions ($\zeta_i=-1$), where the particle $1$ and $2$ are assigned to a fermionic dark matter $\chi$ and its antiparticle $\bar{\chi}$, and the particles $3$ and $4$ are assigned to a standard model fermion $f$ and its antiparticle $\bar{f}$. 
\item The number density of dark matter $\chi$ is only affected by pair annihilation and pair creation processes: $\chi+\bar{\chi} \leftrightarrow f + \bar{f}$.
\item Although the equilibrium distribution does not exactly satisfy the Boltzmann equation, the equilibrium is approximately described by a distribution function with a time-dependent effective chemical potential $\alpha(t)$ \cite{Bernstein1985PRD}:
\begin{eqnarray}
f = \frac{1}{e^{E(p)/T(t) + \alpha(t)} + 1}.
\label{Eq:f_alpha}
\end{eqnarray}
The distribution of antiparticles has $\bar{\alpha}(t)$ instead of $\alpha(t)$.
\item In the beginning, the initial condition $\alpha(t) = \bar{\alpha}(t)=0$ holds \cite{Lee1977PRL}. Since the evolution equations for particles and antiparticles are identical, at a later time, we have $\alpha(t)=\bar{\alpha}(t)$ \cite{Bernstein1985PRD}. 
\item The annihilation products $f$ and $\bar{f}$ are in thermal equilibrium, e.g., $f_f=f_f^{\rm EQ}$ and $f_{\bar{f}}=f_{\bar{f}}^{\rm EQ}$ (or equivalently, $\mu_f=\mu_{\bar{f}}=0$), where $f_f^{\rm EQ}$ and $f_{\bar{f}}^{\rm EQ}$ denote equilibrium distribution functions of $f$ and $\bar{f}$, respectively.
\item The dark matter $\chi$ (and $\bar{\chi}$) approximately obeys the Maxwell-Boltzmann distribution.
\item Since the energies of $f$ and $\bar{f}$ are much greater than the temperature, $f_f\ll 1$, $f_{\bar{f}}\ll 1$, and the Pauli blocking factors are replaced by unity: $(1-f_f)(1-f_{\bar{f}}) = 1$.
\end{enumerate}

Under the assumptions 1 to 6, the Boltzmann equation for fermionic dark matter becomes
\begin{eqnarray}
&& \frac{dn_\chi}{dt} + 3Hn_\chi  \\
&&\quad = -\int d \Pi_\chi d \Pi_{\bar{\chi}} d \Pi_f d \Pi_{\bar{f}} \nonumber \\
&&\qquad \times (2\pi)^4 \delta^4(p_\chi+p_{\bar{\chi}}-p_f-p_{\bar{f}}) \left| \mathcal{M} \right|^2 \nonumber \\ 
&& \qquad \times \left\{  f_\chi f_{\bar{\chi}} (1 - f_f)(1 - f_{\bar{f}}) - f_f f_{\bar{f}} (1 - f_\chi)(1- f_{\bar{\chi}}) \right\} \nonumber
\end{eqnarray}
where
\begin{eqnarray}
\left| \mathcal{M}_{\chi+\bar{\chi} \rightarrow f+\bar{f}} \right|^2 = \left| \mathcal{M}_{f+\bar{f} \rightarrow \chi+\bar{\chi}} \right|^2 = \left| \mathcal{M} \right|^2,
\end{eqnarray}
and
\begin{eqnarray}
&& f_\chi f_{\bar{\chi}} (1 - f_f)(1 - f_{\bar{f}}) - f_f f_{\bar{f}} (1 - f_\chi)(1- f_{\bar{\chi}}) \nonumber\\
&&\quad = f_\chi f_{\bar{\chi}}(1 - f_f^{\rm EQ})(1 - f_{\bar{f}}^{\rm EQ}) ( 1-e^{2\alpha} ) \\
&&\quad = f_\chi^{\rm EQ} f_{\bar{\chi}}^{\rm EQ}e^{-2\alpha} (1 - f_f^{\rm EQ})(1 - f_{\bar{f}}^{\rm EQ})( 1-e^{2\alpha}). \nonumber 
\end{eqnarray}
From the assumption 7, we have
\begin{eqnarray}
n_\chi = e^{-\alpha} n_{\chi}^{\rm EQ},  \quad
n_{\bar{\chi}} = e^{-\alpha} n_{\bar{\chi}}^{\rm EQ}, 
\end{eqnarray}
and
\begin{eqnarray}
e^{-2\alpha} -1 = \left(n_\chi^{\rm EQ} \right)^{-2} \left[  n_\chi^2- \left( n_\chi^{\rm EQ}\right)^2 \right],
\end{eqnarray}
which, finally, leads to the commonly used Boltzmann equation:
\begin{eqnarray}
\frac{dn_\chi}{dt} + 3Hn_\chi = - \langle \sigma v \rangle  \left[  n_\chi^2- \left( n_\chi^{\rm EQ}\right)^2 \right],
\end{eqnarray}
where, from the assumption 8, the thermally averaged cross section can be estimated by
\begin{eqnarray}
\langle \sigma v \rangle &=& \left(n_\chi^{\rm EQ} \right)^{-2} \int d \Pi_\chi d \Pi_{\bar{\chi}} d \Pi_f d \Pi_{\bar{f}} \nonumber \\
&&\quad \times (2\pi)^4 \delta^4(p_\chi+p_{\bar{\chi}}-p_f-p_{\bar{f}}) \left| \mathcal{M} \right|^2 \nonumber\\
&& \quad \times e^{-E_\chi/T} e^{-E_{\bar{\chi}}/T}.
\label{Eq:sigmav}
\end{eqnarray}
%

\subsection{Parafermionic dark matter\label{subsection:parafermionic_dark_matter}}
The Boltzmann equation for fermionic dark matter is extended to include parafermions. Recall that the term of $(1 - f_i)$ shows the Pauli blocking of fermion $i$, which suggests us use $(r_i-f_i)$ as {\it parafermi blocking factor} for a parafermion $i$ with maximum occupation-number $r_i$. The extended Boltzmann equation for $1+2 \leftrightarrow 3+4$ process is 
\begin{widetext}
\begin{eqnarray}
 \frac{dn_1}{dt} + 3Hn_1 &=& -\int d \Pi_1 d \Pi_2 d \Pi_3 d \Pi_4 (2\pi)^4 \delta^4(p_1+p_2-p_3-p_4)  \\ 
&& \quad \times\left\{  \left| \mathcal{M}_{1+2 \rightarrow 3+4} \right|^2 f_1 f_2 (R_3 + Z_3 f_3)(R_4 + Z_4 f_4) -  \left| \mathcal{M}_{3+4 \rightarrow 1+2} \right|^2 f_3 f_4 (R_1 + Z_1 f_1)(R_2 + Z_2 f_2) \right\}, \nonumber
\end{eqnarray}
\end{widetext}
where
\begin{eqnarray}
R_i = 
\begin{cases}
1 & r_i \rightarrow \infty \quad {\rm (bosons)}\\
1 & r_i = 1 \quad {\rm (fermions)} \\
r_i & {\rm other}  \quad {\rm (parafermions)} \\
\end{cases}
\end{eqnarray}
and
\begin{eqnarray}
Z_i = 
\begin{cases}
+1 & r_i \rightarrow \infty \quad {\rm (bosons)}\\
-1 & r_i =1 \quad  {\rm (fermions)}\\
-1 & {\rm other} \quad  {\rm (parafermions)}\\
\end{cases}
\end{eqnarray}
%

\begin{figure}[t]
\begin{center}
\includegraphics{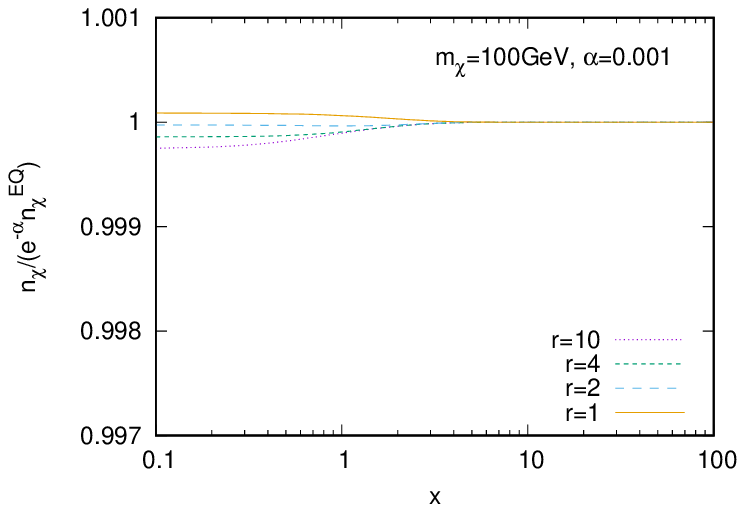}
\includegraphics{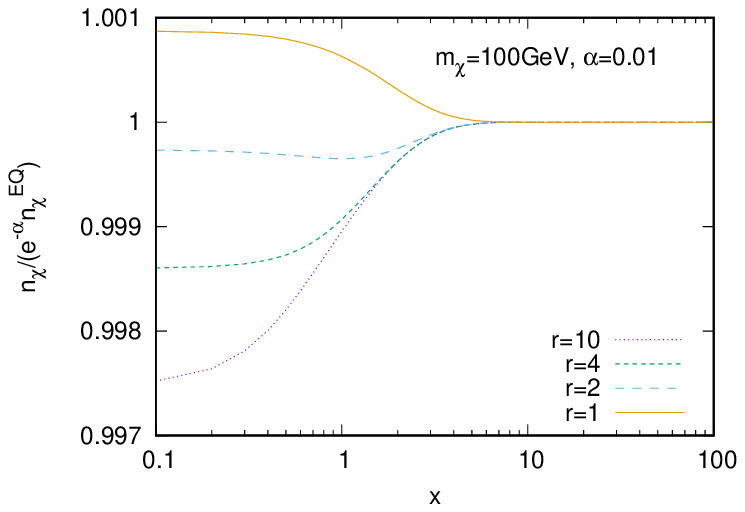}
\caption{The validity of the assumption 4 for parafermions. The upper panel shows the dependence $n_\chi/(e^{-\alpha} n_\chi^{\rm EQ})$ on $x=m_\chi/T$ for $\alpha=0.001$. The lower panel is the same as the upper panel but for $\alpha=0.01$. The anticipated relation of $n_\chi/(e^{-\alpha} n_\chi^{\rm EQ}) \sim 1$ is satisfied.}
\label{fig:n_ratio}
\end{center}
\end{figure}

\begin{figure}[t]
\begin{center}
\includegraphics{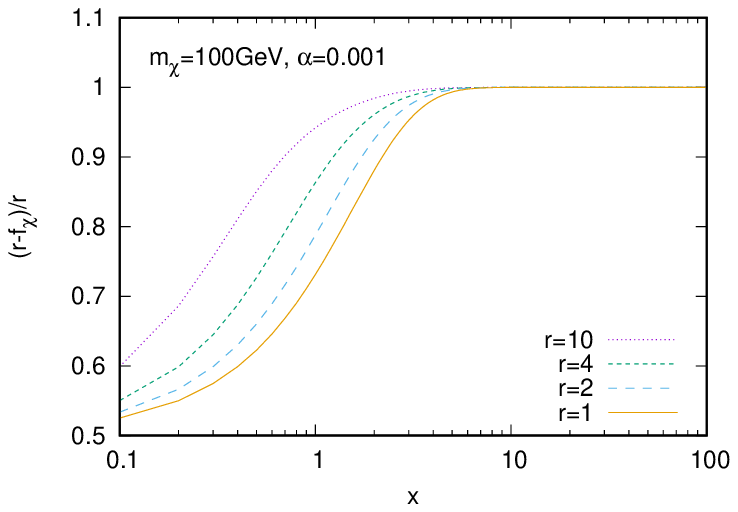}
\includegraphics{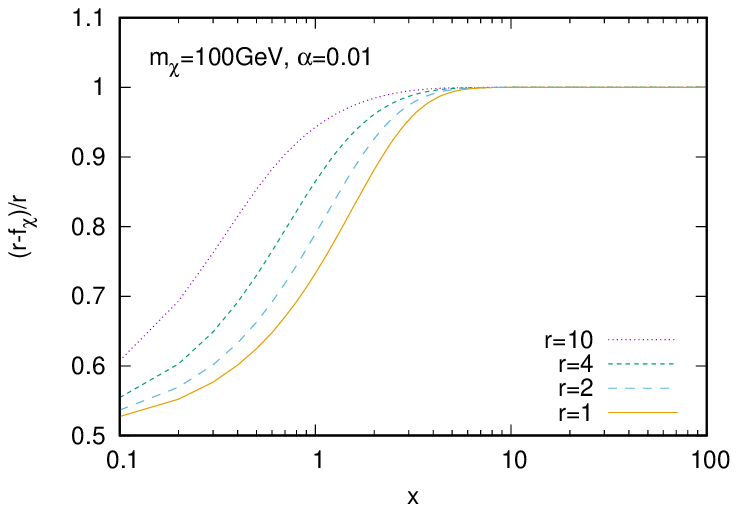}
\caption{The validity of the assumption 5 for parafermions. The upper panel shows the dependence $(r-f_\chi)/r$ on $x=m_\chi/T$ for $\alpha=0.001$. The lower panel is the same as the upper panel but for $\alpha=0.01$. The anticipated relation of $(r-f_\chi)/r \sim 1$ is satisfied for $x \gtrsim 20$.}
\label{fig:r_f}
\end{center}
\end{figure}

Taking into consideration the eight assumptions for fermionic dark matter, we employ the following assumptions for parafermionic dark matter:
\begin{enumerate}
\item The assumptions 1, 3, 5 and 6 for fermionic dark matter can be applied to parafermionic dark matter.
\item The particles $1$ and $2$ are parafermions ($R_i=r_i$, $Z_i=-1$), and $3$ and $4$ are fermions ($R_i=1$, $Z_i=-1$), where the particles $1$ and $2$ serve as parafermionic dark matter $\chi$ with maximum occupation-number $r$ $(2 \le r < \infty)$ and its antiparticle $\bar{\chi}$ and the particles $3$ and $4$ are assigned to a standard model fermion $f$ and its antiparticle $\bar{f}$. 
\item Parafermions are controlled by the distribution function also with a time-dependent effective chemical $\alpha$ defined by 
\begin{eqnarray}
f = \frac{\sum_{n=0}^{r} n e^{-(E/T-\alpha)n}}{\sum_{n=0}^{r} e^{-(E/T-\alpha)n}}.
\end{eqnarray}
\item Recall that $n_\chi = e^{-\alpha} n_{\chi}^{\rm EQ}$ if the fermionic dark matter $\chi$ approximately obeys the Maxwell-Boltzmann distribution.  Similarly, for parafermions, we assume that 
\begin{eqnarray}
f_\chi \sim e^{-\alpha}f_\chi^{\rm EQ},  \quad f_{\bar{\chi}} \sim  e^{-\alpha}f_{\bar{\chi}}^{\rm EQ}, 
\end{eqnarray}
as well as 
\begin{eqnarray}
n_\chi \sim e^{-\alpha} n_\chi^{\rm EQ}, \quad
n_{\bar{\chi}} \sim e^{-\alpha} n_{\bar{\chi}}^{\rm EQ}.
\end{eqnarray}

To see the validity of this assumption, we have performed numerical calculations and the result are shown in FIG.\ref{fig:n_ratio}. The upper panel shows the dependence of the ratio $n_\chi/(e^{-\alpha} n_\chi^{\rm EQ})$ on $x=m_\chi/T$ for $m_\chi=100$ GeV and $\alpha=0.001$ where the number density $n_\chi$ is obtained from Eq.(\ref{Eq:n_rela}). The lower panel is the same as the upper panel but for $\alpha=0.01$. The curve for $r=1$ denotes the ratio for fermion and for $r \ge 2$ denotes the ratio for parafermion with maximum occupation-number $r$. The anticipated relation of $n_\chi/(e^{-\alpha} n_\chi^{\rm EQ}) \sim 1$, as well as $n_\chi \sim e^{-\alpha} n_\chi^{\rm EQ}$, is satisfied without $\alpha$ dependence. We have obtained the similar results for $m_\chi=500$ GeV and $m_\chi=1000$GeV.

\item For fermions, $f_f\ll 1$ and $f_{\bar{f}}\ll 1$ are good approximation, and the Pauli blocking factors are replaced by unity: $(1-f_f)(1-f_{\bar{f}}) = 1$. Similarly, for parafermions, we use $f_\chi\ll 1$ and $f_{\bar{\chi}}\ll 1$ and the parafermi blocking factors replaced by $(r-f_\chi)(r-f_{\bar{\chi}}) \sim r^2$.

FIG.\ref{fig:r_f} shows the validity of this assumption. The upper panel shows the dependence of the ratio $(r-f_\chi)/r$ on $x=m_\chi/T$ for $m_\chi=100$ GeV and $\alpha=0.001$. The lower panel is the same as the upper panel but for $\alpha=0.01$. The anticipated relation of $(r-f_\chi)/r \sim 1$, as well as $r-f_\chi \sim r$, is satisfied without $\alpha$ dependence for $x \gtrsim 20$ (the typical freeze-out temperature $x\sim 25$ \cite{KolbTurner1990}). We have obtained the similar results for $m_\chi=500$ GeV and $m_\chi=1000$GeV.

\end{enumerate}

According to the assumptions 1 to 3, the Boltzmann equation for parafermionic dark matter can be approximated to be
\begin{eqnarray}
&& \frac{dn_\chi}{dt} + 3Hn_\chi  \\
&&\quad = -\int d \Pi_\chi d \Pi_{\bar{\chi}} d \Pi_f d \Pi_{\bar{f}} \nonumber \\
&&\qquad \times (2\pi)^4 \delta^4(p_\chi+p_{\bar{\chi}}-p_f-p_{\bar{f}}) \left| \mathcal{M} \right|^2 \nonumber \\ 
&& \qquad \times \left\{  f_\chi f_{\bar{\chi}} (1 - f_f)(1 - f_{\bar{f}}) - f_f f_{\bar{f}} (r - f_\chi)(r- f_{\bar{\chi}}) \right\} \nonumber
\end{eqnarray}
where
\begin{eqnarray}
&& f_\chi f_{\bar{\chi}} (1 - f_f)(1 - f_{\bar{f}}) - f_f f_{\bar{f}} (r - f_\chi)(r- f_{\bar{\chi}}) \nonumber \\
&&  =  f_\chi^{\rm EQ} f_{\bar{\chi}}^{\rm EQ} \left(e^{-2\alpha} - r^2 \right).
\end{eqnarray}
From the assumption 4 and 5, we can rewrite the factor $e^{-2\alpha} -r^2$ as 
\begin{eqnarray}
e^{-2\alpha} -r^2 = \left(n_\chi^{\rm EQ} \right)^{-2} \left[  n_\chi^2- r^2\left( n_\chi^{\rm EQ}\right)^2 \right],
\end{eqnarray}
and we have the following form of the Boltzmann equation for parefermionic dark matter 
\begin{eqnarray}
\frac{dn_\chi}{dt} + 3Hn_\chi = - \langle \sigma v \rangle  \left[  n_\chi^2- r^2\left(n_\chi^{\rm EQ}\right)^2 \right],
\end{eqnarray}
where the thermally averaged cross section is the same as in Eq.(\ref{Eq:sigmav}). 

Using the standard definitions $Y_\chi=n_\chi/s$ and $x=m_\chi/T$, we obtain
\begin{eqnarray}
\frac{dY_\chi}{dx}  = - \frac{\lambda \langle \sigma v \rangle}{x^2}  \left[ Y_\chi^2- r^2\left( Y_\chi^{\rm EQ}\right)^2 \right] 
\label{Eq:dYdx}
\end{eqnarray}
where
\begin{eqnarray}
\lambda = \frac{4\pi}{\sqrt{90}}M_{pl}m_\chi \sqrt{g_\ast}.
\end{eqnarray}
%

\section{Relic abundance\label{section:relic_abundance}}
\begin{figure}[t]
\begin{center}
\includegraphics{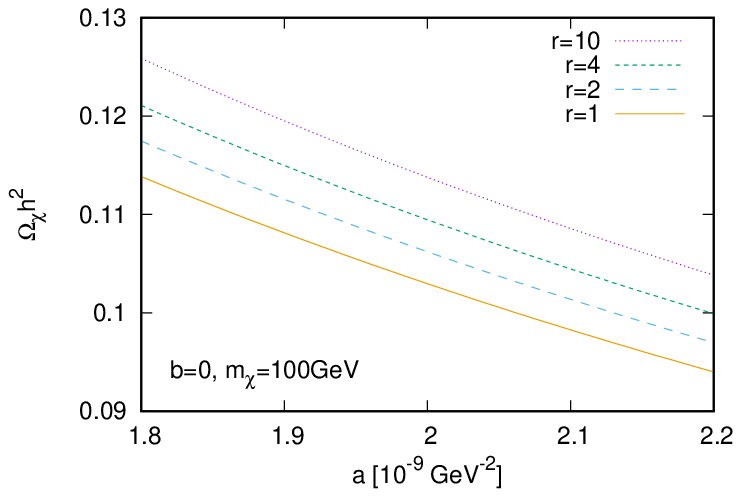}
\includegraphics{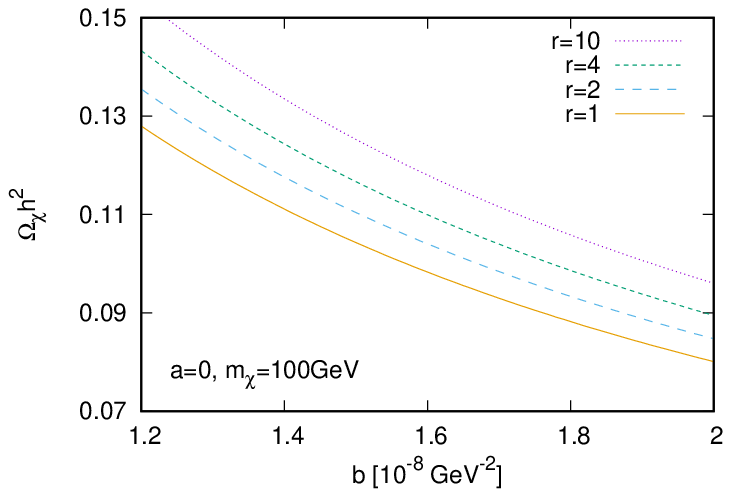}
\caption{Relic abundance of parafermionic dark matter in the s-wave annihilation dominant case (upper) and in the p-wave annihilation dominant case (lower). The parafermionic dark matter ($r \ge 2$) is more abundant in the present universe than the fermionic dark matter ($r=1$).}
\label{fig:omega}
\end{center}
\end{figure}

The present relic abundance of the parafermionic dark matter $\chi$ can be described by the density parameter $\Omega_\chi$ times the scale factor of the Hubble expansion rate $h=0.677$ \cite{Planck2016AA}:
\begin{eqnarray}
\Omega_\chi h^2 = \frac{\rho_\chi}{\rho_{\rm crit}} h^2= \frac{m_\chi s_0 Y_\chi(x\rightarrow \infty)}{\rho_{\rm crit}} h^2,
\label{Eq:Omegah2_chi}
\end{eqnarray}
where $s_0=2.89 \times 10^3$ cm$^{-3}$ and $\rho_{\rm crit} = 3H_0^2/(8\pi G)=1.05h^2 \times 10^{-5}$ GeV cm$^{-3}$ are the present entropy density and the present critical density, respectively, with the Hubble expansion rate $H_0 = 100h$ km s$^{-1}$ Mpc$^{-1}$ and the Newtonian gravitational constant $G=6.67\times 10^{-11}$ m$^3$ kg$^{-1}$ s$^{-2}$ \cite{PDG}. The observed energy density of the cold dark matter component in $\Lambda$CDM model by the Planck Collaboration is $\Omega h^2 = 0.1188 \pm 0.0001$ \cite{Planck2016AA}. 

The existence of relativistic parafermions yields the change of the relativistic degrees of freedom $g_*$ (see Eq.(\ref{Eq:gast}) and FIG.{\ref{fig:gamma} for relativistic parafermions). However, contributions to $g_*$ become negligible, in general, for non-relativistic particles at freeze-out temperature. In this paper, we assume that parafermionic dark matter is cold dark matter consisting of non-relativistic particles, which is described by the standard freeze-out scenario of non-relativistic particles. As a result, we can ignore the effect from the existence of parafermions on $g_*$.

FIG.\ref{fig:omega} shows the relic abundance of parafermionic dark matter. The standard approximation of the averaged cross section
\begin{eqnarray}
\langle \sigma v \rangle = a + 6b x^{-1}
\end{eqnarray}
is employed \cite{Scherrer1986PRD}.  The upper panel shows the dependence of relic abundance $\Omega_\chi h^2$ for s-wave annihilation dominant case ($a\ne 0$ and $b=0$) with $m=100$ GeV and $g_\ast=90$. The lower panel shows the same of the upper panel but for p-wave annihilation dominant case ($a=0$ and $b\ne 0$). 

The parafermionic dark matter ($r \ge 2$) can be more abundant in the present universe than the fermionic dark matter ($r=1$) if the annihilation cross section $\sigma$ does not strongly depend on the maximum occupation-number $r$. For example, the relic abundance is increased by a few \% from fermion $r=1$ to parafermion with maximum occupation-number $r=2$. We have obtained the similar results for $m_\chi=500$ GeV and $m_\chi=1000$GeV.

This increase of relic abundance is qualitatively understood from the Boltzmann equation in Eq.(\ref{Eq:dYdx}). Since the increase of maximum occupation number $r$ yields the increase of the effective initial number density $r Y_\chi^{\rm EQ}$, we observe that the parafermionic dark matter survives longer than fermionic dark matter provided that $\sigma$ is independent of $r$.

\section{summary\label{section:summary}}
We have extended the Boltzmann equation for fermionic dark matter to include parafermions. The parafermion  accommodates $r$ particles per quantum state $(2\le r<\infty)$. The extended Boltzmann equation has the parafermi blocking factor $(r_i-f_i)$ for a parafermion $i$ with maximum occupation-number $r_i$. 

Considering the commonly accepted assumptions for fermionic dark matter, we have derived a simple-formed Boltzmann equation to estimate the relic abundance of parafermionic dark matter. As a result, we find that the parafermionic dark matter can be more abundant than the fermionic dark matter in the present universe. For example, the relic abundance is increased by a few \% from fermion $r=1$ to parafermion with maximum occupation-number $r=2$. 

\vspace{3mm}






\end{document}